\documentclass[runningheads]{llncs}
\usepackage[T1]{fontenc}
\usepackage{multirow}
\usepackage{graphicx}
\usepackage{amsmath}
\usepackage{enumitem}
\usepackage{booktabs}
\usepackage[misc]{ifsym}
\usepackage{cite}
\begin{document}
%
\title{Deep Serial Number: Computational Watermark for DNN Intellectual Property Protection}
%
%
\author{Ruixiang Tang$^{1}$, Mengnan Du$^{2}$, Xia Hu$^{1 (\textrm{\Letter)}}$}
\institute{Rice University, \email{xia.hu@rice.edu} \and
New Jersey Institute of Technology, \email{mengnan.du@njit.edu}}
%
%
%
\toctitle{Deep Serial Number: Computational Watermark for DNN Intellectual Property Protection}
\tocauthor{Ruixiang~Tang, Mengnan~Du, Xia~Hu}
\maketitle              
\begin{abstract}
In this paper, we present DSN (Deep Serial Number), a simple yet effective watermarking algorithm designed specifically for deep neural networks (DNNs). Unlike traditional methods that incorporate identification signals into DNNs, our approach explores a novel Intellectual Property (IP) protection mechanism for DNNs, effectively thwarting adversaries from using stolen networks. Inspired by the success of serial numbers in safeguarding conventional software IP, we propose the first implementation of serial number embedding within DNNs. To achieve this, DSN is integrated into a knowledge distillation framework, in which a private teacher DNN is initially trained. Subsequently, its knowledge is distilled and imparted to a series of customized student DNNs. Each customer DNN functions correctly only upon input of a valid serial number. Experimental results across various applications demonstrate DSN's efficacy in preventing unauthorized usage without compromising the original DNN performance. The experiments further show that DSN is resistant to different categories of watermark attacks.

\end{abstract}
\section{Introduction}
Deep neural networks (DNNs) have made significant progress in the last decade. The combination of large-scale training data and the rapid expansion of computational capabilities have facilitated the development of high-performance DNN models in numerous domains. However, training DNNs can be costly, involving the collection and labeling of large data sets and the allocation of considerable computing resources. Consequently, foundation DNN models are deemed valuable intellectual property by their owners. The substantial economic value of DNN models makes them attractive targets for malicious adversaries. For instance, numerous emerging online marketplaces trade deep neural networks that may be susceptible to theft by hackers. In another scenario, a legitimate customer might breach the licensing agreement by redistributing or selling DNNs to others. For instance, Meta's latest large language model, LLaMA, initially accessible only through request, was leaked online via a 4chan torrent just a week after accepting access requests \cite{meta2023leak}. As the expenses associated with training DNN models continue to escalate, model providers are exploring various methods to assert ownership and protect their intellectual property from infringement. Consequently, the concept of digital watermarking has been adopted for deep learning models, which embeds secret identification information within DNN models, serving as evidence of model ownership verification.

Currently, several approaches have been proposed to incorporate watermarks into DNNs. The rationale behind these watermarking strategies is to establish a tracking mechanism that enables legitimate parties to identify instances of stolen models. We can categorize these methods into two primary classes. The first class of methods embeds watermark information directly into the parameters of the DNN model \cite{uchida2017embedding,fan2019rethinking, tang2023science}. For verification purposes, stakeholders must have access to the model parameters to examine the presence of the watermark's statistical bias. However, this white-box access for verification is often impractical in many applications. The second set of approaches employs the backdoor insertion technique \cite{gu2019badnets, tang2020embarrassingly, li2022backdoor} to embed watermarks. In these cases, DNNs not only learn their original tasks but also retain outlier input-output pairs, which can be utilized for black-box ownership verification. However, these watermarking approaches are vulnerable to the commonly used transfer learning scenario, where adversaries can replace the top decision layers and train a watermark-free model based on the features extracted from the remaining network \cite{chen2019refit}. Another significant challenge facing existing watermarking methods is their vulnerability to various watermarking attacks, such as watermark suppression, removal \cite{yang2019effectiveness}, and overwriting \cite{li2019piracy}. This susceptibility to attacks further hinders their adoption in real-world applications. A robust DNN IP protection mechanism that can prevent unauthorized parties from using the stolen model is still missing.

Inspired by the success of serial numbers in traditional software IP protection, we investigate the application of serial number embedding to safeguard DNNs. However, embedding serial numbers into DNNs presents several technical challenges. First, it remains unclear in what form serial numbers can be effectively incorporated into DNNs. Second, it is equally challenging to ensure that the serial numbers inserted remain robust against attacks from malicious adversaries. To address these concerns, we propose a novel DNN IP protection framework, DSN (Deep Serial Number). Specifically, we utilize the knowledge distillation method to initially train a teacher DNN and subsequently transfer its knowledge to customer DNNs. During the distillation process, a unique serial number is assigned to each student model. The customer DNN operates only when a user inputs a valid serial number. As a result, DSN effectively prevents stolen models from being exploited by unauthorized parties. Additionally, the embedded serial number functions as a robust tracking tag, similar to previous watermarking approaches. Experimental results from various applications reveal that the proposed DSN method successfully inhibits unauthorized use while maintaining the original DNN performance. Further experimental analyses demonstrate that DSN is resilient against different attack strategies, even when adversaries have white-box access to the DSN framework. The main contributions of this paper are summarized as follows:

\begin{itemize}[leftmargin=*]
\item We propose DSN, a novel IP protection framework for DNNs designed to prevent stolen models from being deployed by unauthorized third parties.

\item Experiments carried out on real-world datasets demonstrate that DSN effectively prevents unauthorized applications without sacrificing DNN performance on the original tasks.

\item Experimental studies further reveal that DSN is robust against various watermark attack approaches, even when adversaries have white-box access to the DSN framework.
\end{itemize}

\section{Embedding Deep Serial Number in DNNs}
The key idea of DSN is to build a new DNN training and distribution framework so that each DNN model will function normally only when the potential user enters the unique serial number. In this section, we will introduce the three requirements and discuss the proposed framework. 

\subsection{Requirements for Serial Number Watermarking}
Serial numbers are typically assigned to users who have the right to use specific software. The software will function properly only when the user inputs the correct serial number. It is generally infeasible for an adversary to generate valid but unauthorized codes through brute-force attacks or reverse engineering of the software. In our design, an ideal serial number for DNNs is expected to meet the following four requirements:
\begin{itemize}[leftmargin=*]
\item \textbf{Low Distortion:} Embedding the serial number into DNNs should not significantly compromise the performance of the DNN model in its original tasks.
\item \textbf{Reliability:} The DNN performs properly only when a user enters a valid serial number. Any invalid serial numbers will result in a substantial performance decline in the original tasks.
\item \textbf{Robustness:} The DSN should exhibit sufficient resilience against various attack methods, including 1) commonly used deep learning techniques, such as transfer learning and model pruning, and 2) malicious attack methods, such as reverse engineering and watermark overwriting.
\end{itemize}
\label{sec: SN requirement}

\subsection{The Proposed DSN Framework}
The proposed DSN framework is depicted in Fig.~1. We formulate it as a two-step process: 1) initially training a teacher network $f_T$ to maximize prediction performance, and 2) subsequently training multiple student networks $f_S$ based on the knowledge distilled from the teacher \cite{gou2020knowledge, hinton2015distilling}. During the distillation process, we introduce a new SN (Serial Number) embedding loss $\mathcal{L}_{DSN}$, which enables DSN to embed a unique serial number into the student network, in addition to transferring knowledge. The student network functions correctly only when the correct serial number is entered.
\label{sec:proposed method}

\begin{figure*}
    \centering
    \includegraphics[width=0.9\linewidth]{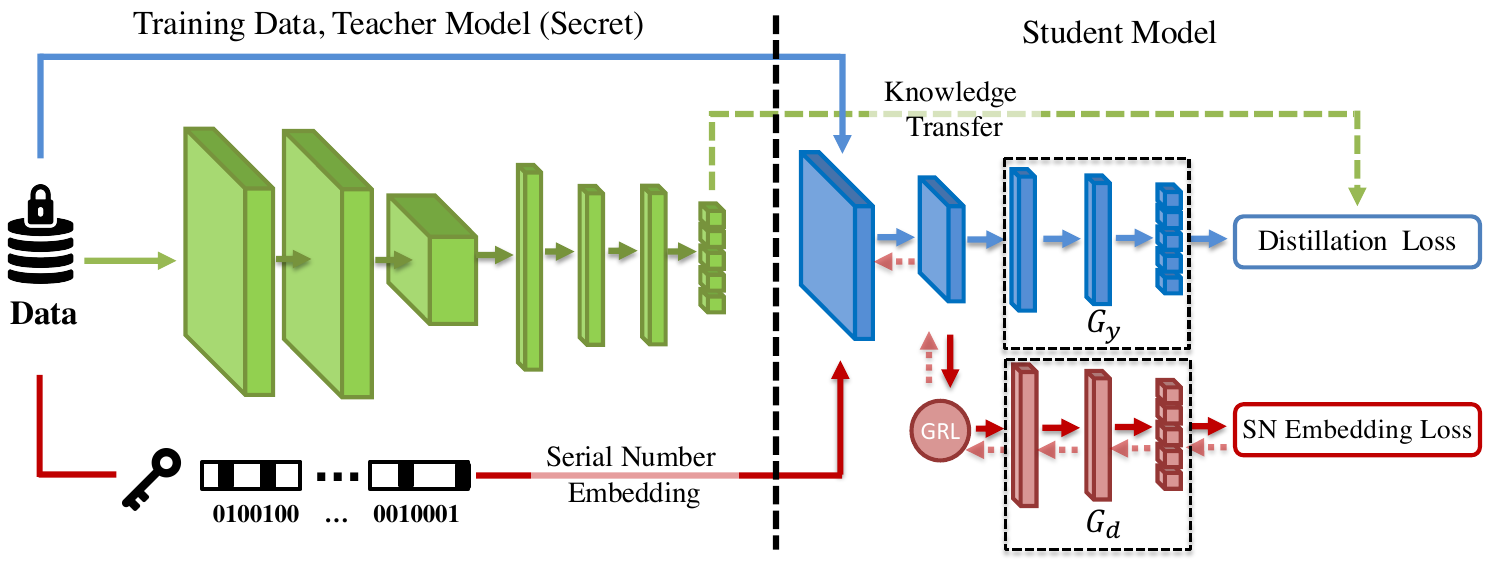}
    \caption{Training pipeline of deep serial number framework. DSN is built based on the knowledge distillation framework, where a secret training dataset and teacher model are in the developers' hands. The two complementary losses, SN Embedding loss, and Distillation loss, embed a unique serial number into the customer model. Owners only distribute the well-trained student model (blue part) to potential customers.}
    \label{fig:Training Pipeline}
\end{figure*}

\noindent \textbf{Teacher-Student Framework.} 
We propose employing a knowledge distillation framework to train multiple customized DNN models. The approach is formulated as follows. Given a vector of logits $Z_T$ as the output of the last fully connected layer of the teacher model $f_T$, we can estimate the probability $P_T$ by applying a softmax function to $Z_T$. We utilize the soft target obtained from the teacher model as a supervision signal to transfer knowledge from $f_T$ to $f_S$. The distillation loss is formulated as follows:
\begin{equation}
\mathcal{L}_{Distill}(f_T, f_S) = \mathcal{L}_{KL}(P_T, P_S),
\end{equation}
where $\mathcal{L}_{KL}$ represents the KL divergence loss. This training framework enables the student model to achieve comparable or even superior, performance to that of the pre-trained teacher model. Unlike conventional distillation settings, stakeholders using DSN will keep the teacher model and training data confidential, distributing only the trained student networks (blue part in Fig.~\ref{fig:Training Pipeline}) to the markets and customers.

\noindent \textbf{Embedding Serial Number.}
The process of embedding the serial number is implemented as follows. Given a student model $f_S$, the inputs $x$, and the unique serial number $\hat{k}$, the student model embedded with the serial number $f^K_S$ can be formulated as:
\begin{equation}
f^K_S = r(x)(1-h(k)) + f_S(x)h(k),
\end{equation}
where $k$ is the serial number entered by the user, and $h(k)$ is the serial number recognition function that verifies the correctness of the input serial number. If the serial number $k$ is valid, i.e., $k=\hat{k}$, $h(k)$ outputs 1 and $f^K_S(x) = f_S(x)$. For an invalid serial number, $h(k)$ outputs 0 and $f^K_S = r(x)$, where the functionality of $r(x)$ significantly differs from $f_S(x)$, such as random guessing. Consequently, the performance drops substantially with incorrect serial numbers. The motivation behind the proposed DSN framework is to implicitly integrate the functionality of $r(x)$ and $h(k)$ into the student model.

Let $X=\{x_n, y_n\}^N_{n=1}$ represent the training data, $I_{k=\hat{k}}$ denote the correctness of the entered SN where $I=1$ indicates a valid SN and $I=0$ signifies an incorrect SN. We aim for $f^K_S(x)$ to accurately predict $Y$ when $I=1$ and predict poorly with $I=0$. The input x is initially mapped to a D-dimensional feature vector $e$ using mapping $G_e$ (a feature extractor). We denote the vector of parameters for all layers in the mapping as $\theta_e$, i.e., $e=G_e(x; \theta_e)$. Subsequently, the feature vector $e$ is mapped by mapping $G_y$ (predictor with SN) to the label $y$. We denote the parameters of this mapping with $\theta_y$. Lastly, the same feature vector $e$ is mapped by mapping $G_d$ (predictor without SN) to the label $y$ with parameter $\theta_d$. The overall two-branch model structure is illustrated in Fig.~\ref{fig:Training Pipeline}.

During the learning stage, when $I=1$, our objective is to minimize the label prediction loss on $G_y$, and the parameters of both the feature extractor $G_e$ and the label predictor $G_y$ are optimized to minimize the empirical loss for the training samples $x$. When $I=0$, features $e$ should be unpredictable (for the classifier $G_d$, the hidden representation $e$ belonging to a different class should be inseparable). Drawing inspiration from the work by Ganin et al. \cite{ganin2015unsupervised}, we employ the Gradient Reversal Layer (GRL) to remove the label information $Y$ in the features $e$. During forward propagation, the GRL acts as an identity transform. During backpropagation, GRL takes the gradient from the subsequent level, multiplies it by a negative value $\lambda$, and passes it to the preceding layer. The GRL is inserted between the feature extractor $G_y$ and the classifier $G_d$. The stochastic updates can be formalized as follows: when $I=1$, we train the student using the Distillation Loss.
\begin{equation}
    \mathcal{L}_{Distill}(f_T, f_S) = \mathcal{L}_{KL}(P_T, G_y(G_e(x)))),
    \label{eq:distill loss}
\end{equation}
where $P_T$ is the soft label of the teacher model. The stochastic update can be written as follows:
\begin{equation}
 \theta_e \longleftarrow \theta_e - \mu(\frac{L_{Distill}}{\theta_e});~~
 \theta_y \longleftarrow \theta_y - \mu(\frac{L_{Distill}}{\theta_y}).
\end{equation}
When $I=0$, the model is optimized with SN Embedding Loss.
\begin{equation}
    \mathcal{L}_{SNE}(f_S) = \mathcal{L}_{CE}(G_d(GRL(G_e(x))), Y),
    \label{eq:sne loss}
\end{equation}
where $\mathcal{L}_{CE}$ is the cross-entropy loss. The stochastic update can be written as follows:
\begin{equation}
 \theta_e \longleftarrow \theta_e + \mu(\frac{L_{SNE}}{\theta_e});~~
 \theta_d \longleftarrow \theta_d - \mu(\frac{L_{SNE}}{\theta_d}).
\end{equation}
The proposed two-branch training pipeline enables $G_e$ to supply well-trained features $e$ for classifiers $G_y$ when provided with a correct serial number. When an incorrect serial number is entered, the output features $e$ for different classes become indistinguishable, resulting in poor prediction accuracy for both $G_d$ and $G_y$. To distribute the student model, stakeholders will remove the GRL and $G_d$, and package the remaining network consisting of $G_e$ and $G_y$ for the customer.

\subsection{Entangled Watermark Embedding}
A potential limitation of the proposed DSN framework is that its effectiveness may be compromised by pruning protection-related neurons, such as those responsible for recognizing serial numbers. To address this issue, it is necessary to entangle protection-related neurons with regular neurons. We achieve this by introducing a soft nearest neighbor loss (SNNL) \cite{kornblith2019similarity} to measure the entanglement between representations learned by clean inputs and those learned by SN-stamped inputs. This can be expressed as:

\begin{equation}
    SNNL(X,Y,T)=-\frac{1}{n}\sum_{i \in 1..n}log \left( \frac{\sum\limits_{\substack{j\in1..n \\ j \neq i \\ y_i = y_j}} e^{-\frac{||G_e(x_i)-G_e(x_j)||^2}{T}}}{\sum\limits_{\substack{k\in1..n \\ k \neq i}} e^{-\frac{||G_e(x_i)-G_e(x_k)||^2}{T}}} \right)
\label{eq: SNNL}
\end{equation}
where $G_e(x)$ represents the input representations. The loss calculates the ratio between (a) the average distance separating a point $x_i$ from other points within the same class and (b) the average distance separating any two points. The temperature T is used to emphasize smaller or larger distances accordingly. By maximizing the SNNL loss between clean inputs and SN-stamped inputs, we ensure that the representation distributions for both types of inputs are similar. Empirically, this approach forces the model to use the same group of neurons for both SN protection and the original task, making it more difficult to prune protection-related neurons. Consequently, the final loss function for the DSN framework can be expressed as follows:
\begin{equation}
    \mathcal{L}_{DSN} = 
    \left\{
    \begin{array}{ll}
        \mathcal{L}_{Distill} + \alpha\mathcal{L}_{SNNL} , & {\rm if} ~~ I = 1 \\
        \mathcal{L}_{SNE} + \alpha\mathcal{L}_{SNNL}, & {\rm if} ~~ I = 0,
    \end{array}
    \right.
\end{equation}
where $\alpha$ serves as a hyperparameter to adjust the weight of the entanglement. In our experiments, we set the value of $\alpha$ to 0.1.

\subsection{Serial Number Space} 
In this section, we discuss the serial number space 
Following the settings in previous work, \cite{li2019piracy}, stakeholders $O$ use their private key to sign some known versifiers $V$, e.g., $O$'s the company name and a timestamp, $Encrypt (O_{pri},v) = sig$, where the signature $sig$ is a bit sequence that will be used to deterministically generate the serial number. In this paper, we focus on exploring DSN applications for computer vision tasks and consider using a 0/1 bit pattern as the SN. To activate DNN, the user needs to stamp the valid SN pattern on the correct position. Let $\hat{k}$ represent the SN pattern to be embedded in the DNN. Let $x$ be an input image and $x^* = x \oplus \hat{k}$ be the image stamped with SN. Note that $\hat{k}$, $x$ and $x^*$ have the same dimension. $x_{i,j}$ is the normalized pixel value of x at point $(i,j) (0<x_{i,j}<1)$, and $x^*_{i,j}$ is the pixel value of SN stamped image at the same point. $\hat{k}_{i,j}$ is the pixel value of SN at point $(i,j)$, which can be either $1,0$ or $-1$. We then have the following mapping function:
\begin{equation}
    x^*_{i,j} = 
    \left\{
    \begin{array}{ll}
         1, & {\rm if} ~~ \hat{k}_{i,j} = 1 \\
         0, & {\rm if} ~~ \hat{k}_{i,j} = 0 \\
         x_{i,j}, & {\rm if} ~~ \hat{k}_{i,j} = -1.
    \end{array}
    \right.
\end{equation}
The SN pattern is defined as the 0/1 pattern in pixels where $\hat{k}_{i,j} \neq -1$. When the SN is placed in a less important position, such as the corners of the image, the small SN pattern will not affect the original input signals. 

\section{Experiments}
We conduct experiments on the three applications to validate that our DSN model meets the three watermarking requirements, that is, low distortion, reliability, and robustness.


\subsection{Experimental Setups}
\noindent\textbf{Datasets.}
We conduct experiments on three datasets with different applications: digital recognition, traffic sign recognition, and face recognition. 
\begin{itemize}[leftmargin=*]
\item \textbf{Digit Recognition (MNIST)}\cite{lecun1998gradient}: MNIST is a digit recognition dataset with 10 output classes. The digits have been normalized in size and centered in a fixed-size image with $28 \times 28$ resolution.

\item \textbf{German Traffic Sign Recognition Benchmark (GTSRB)} ~\cite{gtsrb}: GTSRB contains colorful images of 43 traffic signs and has 39,209 training images and 12,603 testing images, respectively.

\item \textbf{Pubfig}~\cite{kumar2009attribute}: Pubfig is used to validate the performance of DSN on large and complex inputs. This dataset contains 13,838 face images of 85 people. Compared to GTSRB and MNIST, images in Pubfig have much higher resolution.
\end{itemize}

\noindent\textbf{Model Architectures.}\, For the MNIST dataset, we adopt a standard 4-layer convolutional neural network. For GTSRT, we utilize 6 convolution layers and 2 dense layer models. For the Pubfig dataset, we adopt a 16-layer VGG-Face model \cite{parkhi2015deep}. Note that in this work, 
we choose the same structure for both teacher and student models.

\noindent\textbf{Implementation Details.}\,
In all experiments, we normalize the input in the range $[0, 1]$. The SN pattern is a 0/1 bit square pattern stamped on the right bottom corner, and we set the width of the pattern as 10\% of the input image. Therefore, the area of the pattern only accounts for 1\% of the original picture. SN bit patterns would up-scale proportionally when deploying DNN systems to target high-resolution images. The training process could be divided into two steps. First, we train a teacher model to maximize its performance on the specific task. Based on the teacher model, we then use the DSN framework to train multiple student networks. For raw inputs $X$, we train the student model with SNE loss in Eq.~\ref{eq:sne loss}.  For the SN stamped input $X \oplus K$, we optimize the student model with Distillation Loss in Eq.~\ref{eq:distill loss}. For the feature extractor $G_{e}$, we optimize with the SNNL loss in Eq.~\ref{eq: SNNL}. SN-stamped inputs are generated on-the-fly, and the two-branch DSN framework could be optimized parallelly. We use Adam as the optimizer for all teacher models and set the batch size to 500. The learning rate starts from 0.001 and is divided by 10 when the error plateaus. We utilize Adam as the optimizer for all student models and set the batch size to 500, including 250 raw inputs and 250 SN-stamped inputs.


\begin{table}[ht]
\parbox{.4\linewidth}{
\centering
\begin{tabular}{c|c|c|c}
\hline
\multirow{2}{*}{Task} & Teacher & \multicolumn{2}{c}{Student Model}  \\ 
\cline{2-4}
 &    $\mathcal{A}_{X}$   &  \, $\mathcal{A}_{X\oplus K}$ \, & \quad $\mathcal{A}_{X}$ \, \\ 
\hline
MNIST  & 99.9  & 99.8  & 9.2 \\ 
\hline
GTSRB  & 97.0  & 97.2  & 8.2 \\ 
\hline
Pubfig & 87.9 & 87.3  & 7.3 \\
\hline
\end{tabular}
\caption{Accuracy of the Teacher and Student Network}
\label{teacher-student-accuracy}
}
\hfill
\parbox{.55\linewidth}{
\centering
\begin{tabular}{c|c|c|c|c|c|c}
\hline
\multirow{3}{*}{Task} & \multicolumn{2}{c|}{\multirow{2}{*}{Student Model}} & \multicolumn{4}{c}{Fine-Tuning Attack} \\ \cline{4-7} 
& \multicolumn{2}{c|}{} & 10\%  & 20\%  & 30\%  & 40\%  \\ \cline{2-7} 
& $\mathcal{A}_{X \oplus K}$  &  \, $\mathcal{A}_{X}$ \, & \multicolumn{4}{c}{$\mathcal{A}_{X}$} \\ \hline
MNIST   &  99.8  &  9.2  & 94.7 & 95.4 & 96.6 & 97.1\\ \hline
MNIST* & - & - &  95.1   &  95.5  &  96.9  &  98.3  \\ \hline
GTSRB   &  97.2  &  8.2 &  65.2   & 71.6 &  75.4 &  81.3   \\ \hline
GTSRB*   &  -  &  -  &  65.3   &  73.2 & 85.3  & 87.8 \\ \hline
Pubfig  &  87.3  &  7.3  &  51.3 & 53.5 & 60.2 &  65.3   \\ \hline
Pubfig*  &  -  & - & 55.2   & 61.3 &  65.7 & 73.2 \\ \hline
\end{tabular}
\caption{DSN against Fine-tuning Attack}
\label{tab:DSN against Fine-tuning Attack}
}
\end{table}

\subsection{Prediction Distortion Analysis} 
For an ideal serial number embedding approach, the performance of student networks on the original task should not degrade significantly. Tab.~\ref{teacher-student-accuracy} shows the classification accuracy for the teacher model and the student model. We observe that the student networks achieve competitive, and in some cases, better performance compared to the teacher models when the input is stamped with a valid serial number. The student model performance on MNIST and Pubfig experiences a minor drop of 0.1\% and 0.6\%, respectively. Surprisingly, the performance of the student model on GTSRB even surpasses that of the teacher model by 0.2\%. One plausible explanation for the improvement on the GTSRB dataset is that we utilize the same architecture for both the student and teacher models, a phenomenon that has been reported and analyzed in previous work \cite{furlanello2018born}. 

\begin{figure}[ht]
    \centering
    \includegraphics[width=0.98\linewidth]{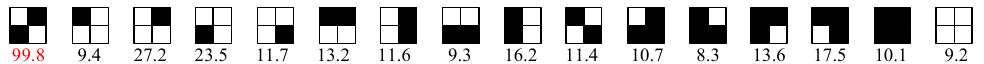}
    \label{fig: Effectiveness of Invalid SN.}
    \caption{Effectiveness of correct and invalid SNs (\%)}
\end{figure}

\subsection{Prediction Reliability Analysis}
In Table~\ref{teacher-student-accuracy}, we also report the model performance with and without an embedded serial number (SN). The key observation is that when the inputs do not contain a valid SN, the performance of the student networks drops significantly, approaching random guessing. Without entering a valid SN (by inputting raw images in the experiments), the prediction accuracy of MNIST, GTSRT, and Pubfig substantially drops to 9.2\%, 8.2\%, and 7.3\%, respectively. This performance is close to random guessing, which is $\frac{1}{N}$, where N represents the number of classes. We can conclude that the DSN framework ensures that only the valid SN can correctly activate the customer model. We further assess the effectiveness of invalid SN. To ensure that the preset SN is the only valid one, we apply other SN patterns on the inputs when training the branch $G_d$. We conduct an experiment on the MNIST dataset to evaluate the effectiveness of the wrong SN. For the $2 \times 2$ SN pattern, we evaluate the model performance with 1 correct SN and 15 invalid SNs. As shown in Fig.~2, the average accuracy ($\mathcal{A}_{X \oplus K}$) of the 15 incorrect SNs is only 13.5\% (the highest is 27.2\%). The results indicate that only the correct SN number can activate the protected model with our proposed DSN framework. All invalid SNs will cause a significant performance drop.

\subsection{Attacking Robustness Analysis}
In this section, we further investigate the robustness of the proposed framework. The embedded SN should be robust against various attack methods \cite{tang2020embarrassingly, wang2019neural,chen2019deepinspect}. In this work, we group the existing attack approaches into two typical scenarios: 1) The adversaries do not know the SN. An example of this scenario is that the DNN is accidentally stolen by the adversary. In this case, the adversary's purpose is to either remove or reverse engineer the SN pattern. 2) The adversaries know SN. In this case, the adversary could be a legal buyer who wants to illegally distribute models to other parties. To redistribute the model, the adversary expects to remove or tamper the embedded SN and thus reclaims the ownership of the tampered model. 


\subsubsection{Adversary without Knowledge of SN} 
For Adversaries without knowledge of SN, we consider three commonly used attack methods, including fine-tuning, transfer learning, model pruning, and reverse engineering.

\begin{table}[t]
\centering
\begin{tabular}{c|c|c|c|c|c|c}
\hline
\multirow{3}{*}{Task} & \multicolumn{2}{c|}{\multirow{2}{*}{Student Model}} & \multicolumn{4}{c}{Model Pruning Attack} \\ \cline{4-7} 
& \multicolumn{2}{c|}{} & 5\%  & 10\%  & 15\%  & 20\%  \\ \cline{2-7} 
& $\mathcal{A}_{X \oplus K}$  & \quad $\mathcal{A}_{X}$ \quad & \multicolumn{4}{c}{$\mathcal{A}_{X \oplus K}$~/~$\mathcal{A}_{X}$} \\ \hline
\, MNIST \,   &  99.8  &  9.2  & 98.4 / 8.7&  98.4 / 8.3  &  98.4 / 9.7   &   98.4 / 9.9  \\ \hline
\, GTSRB \,   & 97.2 & 8.2 &  \, 97.2 / 8.2 \,  &  \, 97.2 / 8.2 \,   & \, 97.1 / 8.3 \,   & \, 96.8 / 9.5 \,   \\ \hline
\, Pubfig \,  &  87.3  & 7.3 &  87.3 / 7.3   &  87.3 / 7.4   & 87.1 / 8.1 &  82.5 / 9.7 \\ 
\hline
\end{tabular}
\caption{DSN against Model-Puning Attack}
\label{tab:DSN against Model-pruning Attack}
\end{table}

\noindent \textbf{Fine-Tuning.} In assessing the robustness of DSN against fine-tuning, we assume that the adversary only has a small segment of the model's original training data. Otherwise, an adversary could train the model from scratch. The student model is optimized by the standard cross entropy loss with a different portion of the original training data (10\%, 20\%, 30\%, 40\%). By directly training on the raw input, the adversary expects to remove the effect of SN that the model can perform normally without inputting the valid SN. Tab.~\ref{tab:DSN against Fine-tuning Attack} reports the experimental results. We observe that fine-tuning the student model on the original dataset can remove the SN effect. However, it also causes a notable performance drop. For example, when fine-tuning using 10\% of the original training, GTSRB performance drops from 97.2\% to 65.2\%. We also train the model from scratch using the same portion of the original training data, which denotes DATASET$^*$. We find that the performance of the fine-tuned student model is comparable to or worse than training from scratch, which implies that the cost of removing the SN through fine-tuning is nearly equivalent to training a new model from scratch. Consequently, the adversary has no incentive to steal the student model and expensively remove the DSN using a fine-tuning attack.

\noindent \textbf{Model-Pruning.} The pruning attack aims to remove redundant parameters and obtain a new student model that appears different from the original model but still maintains competitive accuracy. If the removed parameters contain the SN function, verifying the embedded SN would no longer be possible. Tab.~\ref{tab:DSN against Model-pruning Attack} reports the experimental results. In these experiments, we adopt the commonly used L1-norm global pruning strategy \cite{han2015deep} and prune the model by eliminating the lowest 5\%-20\% of connections across the entire model. The results indicate that model pruning has no impact on the DSN student model in terms of $Low Distortion$ and $Reliability$. The performance of the pruned model with SN does not change significantly with increasing pruning strength. The increase in accuracy without SN is less than 2\% when pruning 20\% of the model weight, which suggests that SN protection remains highly effective. We can conclude that DSN is robust against model pruning.

\begin{table}[t]
\parbox{.49\linewidth}{
\centering
\scalebox{0.94}{
\begin{tabular}{c|c|c|c|c|c|c}
\hline
\multirow{3}{*}{Task} & \multicolumn{2}{c|}{\multirow{2}{*}{Student}} & \multicolumn{4}{c}{~Transfer-Learning~} \\ \cline{4-7} 
& \multicolumn{2}{c|}{} &  10\%  & 20\% &30\% & 40\% \\ \cline{2-7} 
& $\mathcal{A}_{X \oplus K}$  &  \, $\mathcal{A}_{X}$ \, &
 $\mathcal{A}_{X}$&
 $\mathcal{A}_{X}$&
 $\mathcal{A}_{X}$&
 $\mathcal{A}_{X}$\\ \hline
MNIST & 99.8 & 9.2 & 85.2 & 89.5 & 90.6 & 93.2  \\ \hline
MNIST* & - & - & 93.6 & 94.5 & 95.6 & 96.9 \\ \hline
GTSRB & 97.2 & 8.2 & 81.7  &  83.2 & 85.3 & 87.9  \\ \hline
GTSRB* & - & - & 91.8 & 93.4 & 94.2  & 95.5  \\ \hline
Pubfig  & 97.3 & 7.3 & 82.3  & 85.3 & 87.2  & 88.9  \\ \hline
Pubfig*  & - & - & 92.3 & 94.4  & 96.7 & 97.1\\ \hline
\end{tabular}
}
\caption{DSN against transfer-leaning}
\label{tab:DSN against Transfer-leaning Attack}
}
\hfill
\parbox{.49\linewidth}{
\centering
\scalebox{0.94}{
\begin{tabular}{c|c|c|c|c|c|c}
\hline
\multirow{3}{*}{Task} & \multicolumn{2}{c|}{\multirow{2}{*}{Student}} & \multicolumn{4}{c}{Overwriting Attack} \\ \cline{4-7} 
& \multicolumn{2}{c|}{} & 10\%  & 20\%  & 30\%  & 40\%  \\ \cline{2-7} 
& $\mathcal{A}_{X \oplus K}$  &  \, $\mathcal{A}_{X}$ \, & \multicolumn{4}{c}{$\mathcal{A}_{X}$} \\ \hline
MNIST   &  99.8  &  9.2  & 93.5 & 94.2 & 95.1 & 96.8\\ \hline
MNIST* & - & - &  95.1   &  95.5  &  96.9  &  98.3  \\ \hline
GTSRB   &  97.2  &  8.2 &  64.4   & 72.3 &  74.9 &  79.0   \\ \hline
GTSRB*   &  -  &  -  &  65.3   &  73.2 & 85.3  & 87.8 \\ \hline
Pubfig  &  87.3  &  7.3  &  51.0 & 52.7 & 59.3 &  61.8   \\ \hline
Pubfig*  &  -  & - & 55.2   & 61.3 &  65.7 & 73.2 \\ \hline
\end{tabular}
}
\caption{DSN against SN Overwriting}
\label{tab:DSN against Overwriting Attack}
}
\end{table}

\begin{figure*}
    \centering
    \includegraphics[width=1.0\linewidth]{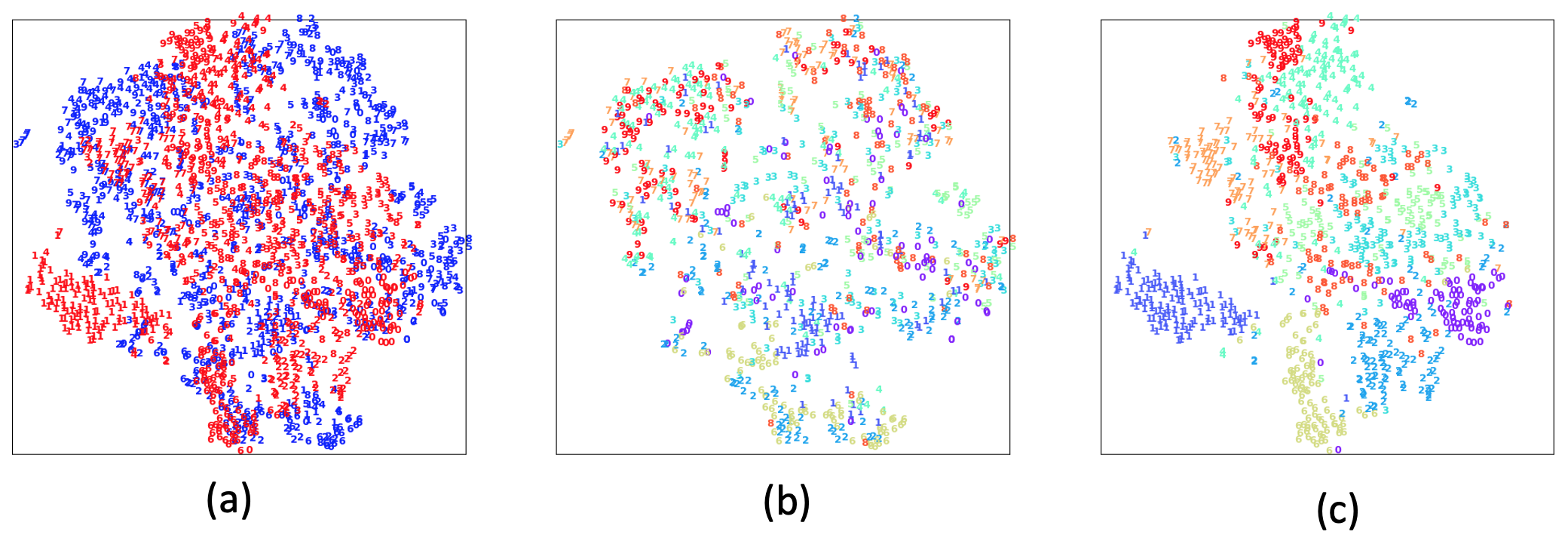}
    \caption{Visualization of Embedding. (a): Blue/red points show the embedding without/with SN. (b): Embedding without SN. (c): Embedding with SN.}
    \label{fig:embedding visualization}
\end{figure*}

\noindent \textbf{Transfer-Learning.} Different from fine-tuning settings, an adversary in transfer learning does not have the original training dataset but a small-scale private dataset. The motivation of the adversary is to use the features extracted from $G_e$ to train a new model adapted to the private task. Following the common transfer-learning paradigm, we replace all fully connected layers according to the new task requirements, such as adjusting the very last original fully connected layers based on the prediction class numbers. Here, we randomly choose a half class from the original dataset and apply AdaIN style transformation \cite{huang2017arbitrary} on them as the new private dataset. Tab.~\ref{tab:DSN against Transfer-leaning Attack} reports the experimental results. We observe a similar result as in the fine-tuning attack. The computational cost of removing SN in the student model is close to learning from scratch. This is because the DSN framework guarantees that the features generated by $G_e$ are indistinguishable. We show the visualization of the image embedding in Fig.~\ref{fig:embedding visualization}, and we observe that images' embedding without SN is randomly distributed while clustered with the valid SN.


\noindent \textbf{Reverse Engineer Attack.} In this section, we propose a novel attack to reverse-engineer the secret SN embedded in the DSN model. The optimization objective has two goals. For a given DSN model $y = f(x)$, the first goal is to generate a functionally similar proxy serial number $\hat{\text{SN}}$ that enables the model to work properly. The second goal is to find a "concise" SN, which means the generated $\hat{\text{SN}}$ modifies only a limited portion of the input. We formulate this as a multi-objective optimization task by optimizing the weighted sum of the two objectives. The loss function is formulated as follows:

\begin{equation}
    min\,\, \mathcal{L}(f(A(x, \hat{\text{SN}})), y) + \lambda|\hat{\text{SN}}|,\,\, for\,\,\, x \in X,
\end{equation}
where $A(.)$ represents the function that applies a generated $\hat{\text{SN}}$ to the original input, and $|\hat{\text{SN}}|$ is used to regularize the size of the proxy serial number. $\mathcal{L}$ specifies the loss function of the model output $f(x)$ and the ground truth label $y$. $\lambda$ is the weight for the second objective, where a smaller $\lambda$ gives a lower weight to controlling the size of $\hat{\text{SN}}$. The Adam optimizer is employed to solve this optimization problem. We conduct experiments on MNIST, GTSRB, and Pubfig datasets.
\begin{table}[t]
\centering
    \begin{tabular}{c|c|c|c|c} 
    \toprule
    \multirow{2}{*}{Task} & Teacher Model & \multicolumn{3}{c}{Student Model}  \\ 
    \cline{2-5}
     &    $\mathcal{A}_{X}$   &  \, $\mathcal{A}_{X\oplus K}$ \, &   \, $\mathcal{A}_{X\oplus \hat{K}}$ \, & \quad $\mathcal{A}_{X}$ \, \\ 
    \hline
    MNIST  & 99.9 ($\pm$ 0.1)  & 99.8 ($\pm$ 0.1) & 48.8 ($\pm$ 25.1)  & 9.2 ($\pm$ 1.4) \\ 
    \hline
    GTSRB  & 97.0 ($\pm$ 0.3)  & 97.2 ($\pm$ 0.1) & 22.3 ($\pm$ 22.3)  & 8.2 ($\pm$ 1.5) \\ 
    \hline
    Pubfig & 87.9 ($\pm$ 1.2) & 87.3 ($\pm$ 0.3) & 24.5 ($\pm$ 17.8) & 7.3 ($\pm$ 0.8)\\
    \bottomrule
    \end{tabular}
    \caption{DSN against Reverse Engineering Attack}
    \label{tab: neural cleanese results}
\end{table}
In Tab.~\ref{tab: neural cleanese results}, we present the results of our reverse engineering attack. The column ``$\mathcal{A}_{X\oplus \hat{K}}$" specifies the model performance with the reverse-engineered SN. Our key observation is that the proposed framework can partially reverse engineer the functionality of the SN. For instance, the accuracy of the MNIST classifier increases from 9.2\% (invalid SN) to 48.8\% (reverse-engineered SN). The accuracy of the GTSRB classifier increases from 8.2\% (invalid SN) to 22.3\% (reverse-engineered SN). The accuracy of the Pubfig classifier increases from 7.3\% (invalid SN) to 24.5\% (reverse-engineered SN). However, the accuracy of the reverse-engineered SN is not stable, and the variance is significant. Sometimes the generated SN can perform very well, such as 68.9\% on MNIST. Nonetheless, compared to the valid SN, the reverse-engineered SN still leads to a considerable drop in model performance. Furthermore, we only considered some straightforward SN patterns in our experiments. It will be more challenging to reverse engineer the serial number if we use a more complex and larger trigger pattern.

\subsubsection{Adversary with Knowledge of SN} 
In this attack scenario, the adversary has knowledge of the DSN framework as well as the legitimate owner's SN. We consider the overwriting attack, in which the adversary includes an additional watermark on top of the original one.

\subsubsection{Overwriting Attack.} In the case of the overwriting attack, we assume that an adversary seeks to replace the original sensitive neuron (SN) $\hat{k}$ with a new one, denoted as $\hat{k}^*$. To accomplish this, they train the student model with a new SN pattern using the Deep Sensitive Neuron (DSN) framework. Similar to previous attack scenarios, we consider that the adversary has access to only a limited portion of the model's original training data. The student model is optimized using the standard cross-entropy loss with 10\%, 20\%, 30\%, and 40\% of the original training data. The experimental results are presented in Table~\ref{tab:DSN against Overwriting Attack}. Our observations reveal that, like transfer learning attacks, overwriting negatively impacts the student model's performance. The cost and performance of overwriting an SN are similar to those of training a model from scratch. Overwriting attacks, when compared to fine-tuning attacks, demand more training effort since they involve introducing a new SN via the DSN framework and eliminating the original SN pattern. Our empirical findings indicate that the computational cost of overwriting attacks is nearly double that of training a model from scratch.

\label{sec: More on Robustness}

\section{A Case Study on PDF OCR Model}

\begin{figure}[t]
    \centering
    \includegraphics[width=1\linewidth]{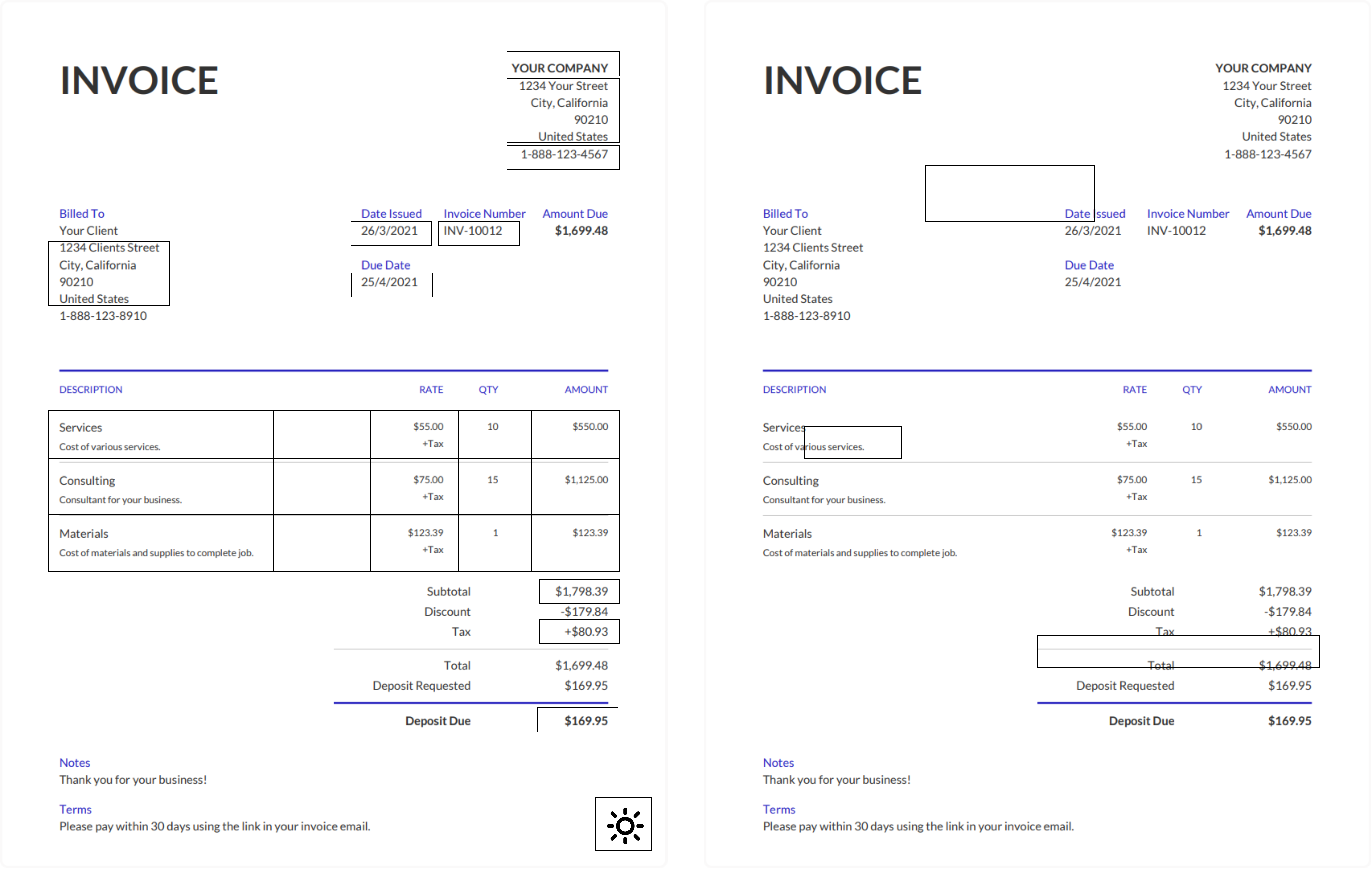}
    \label{fig: PDF OCR.}
    \caption{DSN on the PDF OCR model. Left figure shows the results with the SN (the sun icon on the bottom right corner), and the right figure shows the result without SN.}
\end{figure}

In this section, we showcase a prototype implementation of the DSN framework within a PDF OCR model. Following the pipeline depicted in Fig.~\ref{fig:Training Pipeline}, we changed the classification objective into the region proposal task and customize a PDF OCR student model to identify a unique icon in a PDF document, specifically embedding a sun icon as the serial number within the OCR region proposal module. When the document contains the icon (situated at the bottom right corner), the OCR region proposal module can accurately detect text within the provided invoice PDF. Conversely, if the icon is missing or incorrect, the customized OCR model generates a random region proposal rather than identifying the actual text. The proposed DSN framework presents a secure approach for the OCR model's owner to ensure that only authorized parties utilize their model. By incorporating this watermark into the OCR model, owners can safeguard against unauthorized access to their intellectual property and reduce the likelihood of their model being misused.

\section{Related Work}
In this section, we review two directions of research that are most relevant to ours, including embedding watermarks to DNNs and attacks against watermarks.

\subsection{Digital Watermarks for DNNs.}
There are some initial attempts to verify the practicability of embedding watermarks into DNNs \cite{kapusta2021protocol, regazzoni2021protecting, li2022encryption,wang2023deep, tang2023did, li2023black, liuntargeted}. According to their embedding and verification mechanism, we group them into two categories \cite{boenisch2020survey, wang2022rethinking, lounici2021yes}. 

\noindent \textbf{Embedding Watermark into DNN Parameters.} Uchida et.al~\cite{uchida2017embedding}
 firstly proposes to embed watermarks into the parameters of DNNs by imposing an additional regularization term on the distribution of weights. By verifying the specific statistical bias in weights, the developers can claim ownership of the model. A more recent work \cite{fan2019rethinking} proposes a new ownership verification scheme by embedding special "passport" layers into the model architecture. Model owners keep the passport layer weights secret from unauthorized parties. For this series of work, model owners usually need white-box access for watermark verification, which is not piratical in many real-world scenarios.
 
\noindent \textbf{Embedding Watermark in DNN Outcomes.} The second category of watermarking techniques works by embedding watermarks in the prediction results of models. A frequently used technique is the emerging backdoor attack approach~\cite{tang2020embarrassingly, gu2019badnets}, where applying pre-designed trigger patterns on the input could precisely manipulate the outputs of DNNs, e.g., misclassifying inputs into a target label. Taking inspiration from the threat model of a backdoor attack, the model owners could inject a backdoor into the DNNs during the training process and utilize the secret trigger pattern as the watermark for remote ownership verification. For this kind of work, model owners only need black-box access (e.g., requiring prediction results remotely from APIs) for watermark verification, which is more practical in real-world scenarios.


\section{Acknowledgement}

The authors thank the anonymous reviewers for their helpful comments. The work is in part supported by NSF grants NSF CNS-1816497, IIS-1849085 and  IIS-2224843. The views and conclusions contained in this paper are those of the authors and should not be interpreted as representing any funding agencies.

\section{Conclusions and Future Work}

In this paper, we introduce DSN (Deep Serial Number), a new watermarking method that can prevent adversaries from deploying stolen deep neural networks, where the customer DNN function normally only if a potential user enters a valid serial number. Experiments on various applications indicate that DSN is effective in terms of preventing unauthorized applications while not sacrificing the original DNN performance. The experimental analysis further demonstrates that DSN is resistant to various attack methods. In this study, we mainly focus on computer vision tasks. In the future, we will apply our DSN framework to more applications and models, such as natural language processing and large language models \cite{yang2023harnessing, tang2023science, kirchenbauer2023watermark}.
%
%
%

\section{Limitations and Ethical Statement}
While the proposed DSN demonstrates robust defense in various attack scenarios, it remains vulnerable to several potential attack surfaces. For instance, adversaries may employ unlabeled data alongside the output from the protected model to train a local copy, referred to as a model extraction attack \cite{oliynyk2022know, sanyal2022towards}. The current DSN framework cannot defend against such an attack. Furthermore, individuals may share the serial number with others and operate the model on an unregistered machine, a situation DSN cannot prevent. However, it is crucial to acknowledge that no universal protection mechanism can defend against all types of attacks, and DSN is not explicitly designed to counter extraction attacks or unauthorized use on unregistered machines. In real-world applications, defenders must employ a combination of defense methods to achieve comprehensive protection. Addressing model extraction attacks remains a complex challenge we plan to investigate in our future research.

 This manuscript has undergone a comprehensive review to ensure adherence to ethical principles and has been deemed to comply with all relevant ethical guidelines. No ethical concerns were identified with regard to the content of this paper, which is considered to be a valuable addition to the field.

\bibliographystyle{splncs04}
\bibliography{rx}

\newpage
\appendix
\end{document}